\documentstyle[12pt]{article}
\begin{document}
\input amssym.def
\input amssym
\hfuzz=5.0pt
\def\vec#1{\mathchoice{\mbox{\boldmath$\displaystyle\bf#1$}}
{\mbox{\boldmath$\textstyle\bf#1$}}
{\mbox{\boldmath$\scriptstyle\bf#1$}}
{\mbox{\boldmath$\scriptscriptstyle\bf#1$}}}
\def\mbf#1{{\mathchoice {\hbox{$\rm\textstyle #1$}}
{\hbox{$\rm\textstyle #1$}} {\hbox{$\rm\scriptstyle #1$}}
{\hbox{$\rm\scriptscriptstyle #1$}}}}
\def\operatorname#1{{\mathchoice{\rm #1}{\rm #1}{\rm #1}{\rm #1}}}
\chardef\ii="10
\def\widehat{\mathaccent"0362 }
\def\widetilde{\mathaccent"0365 }
\def\vphi{\varphi}
\def\vrho{\varrho}
\def\vtheta{\vartheta}
\def\ih{{\i\over\hbar}}
\def\CD{{\cal D}}
\def\CL{{\cal L}}
\def\CP{{\cal P}}
\def\CV{{\cal V}}
\def\half{{1\over2}}
\def\bhalf{\hbox{$\half$}}
\def\viert{{1\over4}}
\def\bviert{\hbox{$\viert$}}
\def\oo{\operatorname{o}}
\def\OO{\operatorname{O}}
\def\SO{\operatorname{SO}}
\def\SU{\operatorname{SU}}
\def\ce{\operatorname{ce}}
\def\se{\operatorname{se}}
\def\Mc{\operatorname{Mc}}
\def\Ms{\operatorname{Ms}}
\def\ps{\operatorname{ps}}
\def\dfrac#1#2{\frac{\displaystyle #1}{\displaystyle #2}}
\def\pathint#1{\int\limits_{#1(t')=#1'}^{#1(t'')=#1''}\CD #1(t)}
\def\hbarm{{\dfrac{\hbar^2}{2m}}}
\def\bbbr{{\rm I\!R}} 
\def\bbbn{{\rm I\!N}} 
\def\bbbz{{\mathchoice {\hbox{$\sf\textstyle Z\kern-0.4em Z$}}
{\hbox{$\sf\textstyle Z\kern-0.4em Z$}}
{\hbox{$\sf\scriptstyle Z\kern-0.3em Z$}}
{\hbox{$\sf\scriptscriptstyle Z\kern-0.2em Z$}}}}
\def\d{\operatorname{d}}
\def\e{\operatorname{e}}
\def\i{\operatorname{i}}
 
\thispagestyle{empty}
\centerline{\normalsize DESY 97--197 \hfill ISSN 0418 - 9833}
\centerline{\normalsize October 1997\hfill}
\centerline{\normalsize hep-th/9711118\hfill}
\vskip.3in
\message{TITLE:}
\message{INTEGRABILITY, COORDINATE SYSTEMS, AND SEPARATION OF VARIABLES IN PATH 
         INTEGRALS}
\begin{center}
{\Large INTEGRABILITY, COORDINATE SYSTEMS, AND
\vskip.05in
SEPARATION OF VARIABLES IN PATH INTEGRALS}
\end{center}\vskip.3in
\begin{center}
{\Large Christian Grosche}
\vskip.2in
{\normalsize\em II.\,Institut f\"ur Theoretische Physik}
\vskip.05in
{\normalsize\em Universit\"at Hamburg, Luruper Chaussee 149}
\vskip.05in
{\normalsize\em 22761 Hamburg, Germany}
\end{center}
\vfill
{\small
Contribution to the Proceedings of the {\it VIII.~International Conference on
Symmetry Methods in Physics}, July 28, 1997 -- August 2, 1997, Dubna, Russia.}
\bigskip\bigskip
\begin{center}
{ABSTRACT}
\end{center}
\smallskip
\noindent
{\small
In this contribution I summarize the achievements of separation of variables 
in integrable quantum systems from the point of view of path integrals. This
includes the free motion on homogeneous spaces, and motion subject
to a potential force, and I would like to propose systematic investigations 
of parametric coordinate systems on homogeneous spaces.}

\bigskip
\newpage\noindent
 
 
\normalsize
\noindent
The problem of separation of variables in physical systems has been studied
for a long time. First, the question arises, is a system separable, i.e., is 
it integrable (which is not the same)? Many efforts have been put into the 
investigation to construct separable systems. For the usual physical systems 
this means to consider potential or magnetic forces, breaking the homogeneity 
of the free motion, which still allow complete separation of variables in 
some coordinate system. In three-dimensional space this means that there 
must be at least three functionally independent constants of motion, 
representing the separation constants. If there are less than three constants 
of motion, in a quantum system they are called observables, the physical 
system is not entirely separable, a property which leads usually to some sort 
of chaoticity, e.g.\cite{GUTd}.

If there are more than three observables in a systems, such a system is 
called superintegrable. Generally, in $d$ dimensions a system is called
maximally superintegrable if there are $2d-1$ constants of motion. The best
known examples are the isotropic oscillator and the Coulomb potential in 
spaces of constant curvature, respectively. This property has the consequence 
that separability in more that one coordinate system is possible. For instance,
in three dimensions the oscillator has five constants of motion and is 
separable in eight (out of eleven) coordinate systems, and the Coulomb 
potential has also five constants of motion (including the famous 
Pauli-Lenz-Runge vector) and is separable in four coordinate systems.

Whereas for a particular physical system one is most concerned with
separability, in homogeneous spaces one wants to know the total number of 
separable coordinate systems for the free Schr\"odinger (Helmholtz-) equation, 
or equivalently, how many inequivalent sets of commuting observables can be 
constructed.

Obviously, these two topics attracted a considerable amount of attention.
For the most important two- and three-dimensional spaces of constant curvature,
i.e., Euclidean space, the sphere and the hyperboloid, this question was
settled by Olevski\u\ii\ \cite{OLE} in a classical article. Fischer et al.\
\cite{FNR} studied spherical functions for generalized rotation groups. Later 
on, Kalnins and Miller studied in a series of articles higher dimensional 
spaces, for instance in \cite{KAMIj} the $\OO(2,2)$ hyperboloid, in 
\cite{KAMIo} four-dimensional flat space (Euclidean and Minkowski space); 
systematically, these studied have been summarized by Kalnins \cite{KAL} who 
gave an account how to construct separable coordinate systems in spaces of 
constant curvature in higher dimensions.

Concerning potential problems, Eisenhardt \cite{EIS} investigated the
kind of potentials where the Schr\"odinger equation is soluble, and 
Smorodinski, Winternitz et al.\ \cite{MSVW} started a systematic study of
integrable potentials in three-dimensional Euclidean space.

From the point of view of path integrals one has to deal with two difficulties:
First, the problem is question must be separated, second, expansion formulas
must be found in order to complete the path integration. Whereas it is
straightforward to formulate a separation formula for separable coordinate 
systems \cite{GROad,GRSh,GROPOd}, the finding of expansion formulas is quite
another matter. By an expansion formula we mean, for instance, the expansion
of plane waves into circular waves according to
\begin{equation}
 \e^{z\cos\psi}=\sum_{\nu\in\bbbz}\e^{\i\nu\psi}I_\nu(z)\enspace,
\label{numAb}
\end{equation}
which (and its higher dimensional generalization) lies at the very basis for
path integrals in polar coordinates \cite{GRSb,PI}. Fortunately, more 
expansion formulas have been found and have given rise to set up a set of 
Path Integral Identities which may be called Basis Path Integrals. They are
\begin{enumerate}
\item {\em The Gaussian Path Integral}. This includes the path integral for
      the harmonic oscillator, as well as for the general quadratic Lagrangian.
\item {\em The Besselian Path Integral}. This is the path integral for the
      (generalized) radial harmonic oscillator.
\item {\em The Legendrian Path Integrals}. These two path integrals correspond
      to the path integral solution of the P\"oschl-Teller, respectively
      modified P\"oschl-Teller potential, respectively the Rosen-Morse
      oscillator \cite{BJB,FLMb,KLEMUS}. 
\end{enumerate}

\begin{table}[h]
\caption{\label{cosytab} Path Integration on Homogeneous Spaces}
\begin{eqnarray}\begin{array}{l}\vbox{\small\offinterlineskip
\halign{&\vrule#&$\strut\ \hfil\hbox{#}\hfill\ $\cr
\noalign{\hrule}
height2pt&\omit&&\omit&&\omit&\cr
&Homogeneous Space&&Number of      &&Number of Systems in which &\cr
&   &&Coordinate Systems &&Path Integration is possible&\cr
height2pt&\omit&&\omit&&\omit&\cr
\noalign{\hrule}\noalign{\hrule}
height2pt&\omit&&\omit&&\omit&\cr
&Two-Dimensional  &&               &&                  &\cr
&Pseudo-Euclidean Space  &&10      &&10                &\cr
height2pt&\omit&&\omit&&\omit&\cr
\noalign{\hrule}
height2pt&\omit&&\omit&&\omit&\cr
&Three-Dimensional&&               &&                  &\cr
&Pseudo-Euclidean Space  &&54      &&32                &\cr
height2pt&\omit&&\omit&&\omit&\cr
\noalign{\hrule}
height2pt&\omit&&\omit&&\omit&\cr
&Four-Dimensional&&                &&                  &\cr
&Pseudo-Euclidean Space  &&261     &&182               &\cr
height2pt&\omit&&\omit&&\omit&\cr
\noalign{\hrule}
height2pt&\omit&&\omit&&\omit&\cr
&Two-Dimensional  &&               &&                  &\cr
&Euclidean Space  &&4              &&4                 &\cr
height2pt&\omit&&\omit&&\omit&\cr
\noalign{\hrule}
height2pt&\omit&&\omit&&\omit&\cr
&Three-Dimensional&&               &&                  &\cr
&Euclidean Space  &&11             &&9                &\cr
height2pt&\omit&&\omit&&\omit&\cr
\noalign{\hrule}
height2pt&\omit&&\omit&&\omit&\cr
&Four-Dimensional &&               &&                  &\cr
&Euclidean Space  &&42             &&35                &\cr
height2pt&\omit&&\omit&&\omit&\cr
\noalign{\hrule}
height2pt&\omit&&\omit&&\omit&\cr
&Two-Dimensional  &&               &&                  &\cr
&Sphere           &&2              &&2                 &\cr
height2pt&\omit&&\omit&&\omit&\cr
\noalign{\hrule}
height2pt&\omit&&\omit&&\omit&\cr
&Three-Dimensional&&               &&                  &\cr
&Sphere           &&6              &&5                 &\cr
height2pt&\omit&&\omit&&\omit&\cr
\noalign{\hrule}
height2pt&\omit&&\omit&&\omit&\cr
&Two-Dimensional  &&               &&                  &\cr
&Pseudosphere     &&9              &&9                 &\cr
height2pt&\omit&&\omit&&\omit&\cr
\noalign{\hrule}
height2pt&\omit&&\omit&&\omit&\cr
&Three-Dimensional&&               &&                  &\cr
&Pseudosphere     &&34             &&24                &\cr
height2pt&\omit&&\omit&&\omit&\cr
\noalign{\hrule}
height2pt&\omit&&\omit&&\omit&\cr
&$\OO(2,2)$-Hyperboloid
                  &&72             &&33                &\cr
height2pt&\omit&&\omit&&\omit&\cr
\noalign{\hrule}}}\end{array}\nonumber\end{eqnarray}
\end{table}

\noindent
In \cite{GROad} I have tried to summarize the possibilities of achieving path
integral solutions in homogeneous spaces according to the lines of the Basic
Path Integrals, the results are listed in table \ref{cosytab}. As it turned
out, the above listed Basic Path Integrals are far from complete in doing the 
job. I could find several other expansion formulas and path integral 
identities connected with elliptic and spheroidal coordinates. These expansion 
formulas were used together with interbases expansion relations according to
\begin{equation}
   |\vec k>=\int dE_{\vec l}\,C_{\vec l,\vec k}|\vec l>\enspace.
\label{numAc}
\end{equation}
Here $|\vec k>$ stands for a basis of eigenfunctions of the relevant 
Hamiltonian in the coordinate space representation $\vec k$, and $\int dE_
{\vec l}$ is the spectral-expansion with respect to the coordinate space
representation $\vec l$ with coefficients $C_{\vec l,\vec k}$ which can
be discrete, continuous or both. For example, in two-dimensional Euclidean
space we consider the expansion for an arbitrary $\alpha$ ($h=pd/2$, with $d$ 
the parameter in the elliptic coordinate system $(\mu,\nu)$, and $p$ the 
momentum, \cite[p.185]{MESCH})
\begin{eqnarray}       & &\!\!\!\!\!\!\!\!
   \exp\Big[\i p(x\cos\alpha+y\sin\alpha)\Big]
  =2\sum_{n=0}^\infty\i^n\ce_n(\alpha;h^2)\Mc_n^{(1)}(\mu;h)\ce_n(\nu;h^2)
         \nonumber\\   & &\!\!\!\!\!\!\!\!\qquad\qquad\qquad\qquad
   +2\sum_{n=1}^\infty\i^{-n}
  \se_n(\alpha;h^2)\Ms_{-n}^{(1)}(\mu;h)\se_n(\nu;h^2)\enspace.
\label{numAa}
\end{eqnarray}
$\ce_n,\se_n$, $\Mc_n^{(1)},\Ms_n^{(1)}$ are periodic and non-periodic Mathieu 
functions, which are  even and odd, respectively. Equation (\ref{numAa}) 
represents the expansion of plane waves into elliptical waves, similarly as
the expansion (\ref{numAb}) into circular waves.

The corresponding attempt to review path integrals for separable potentials 
were performed in \cite{GROPOa}--\cite{GROPOd}. However, there the emphasize 
was concentrated on superintegrable systems in spaces of constant curvature, 
i.e., in two- and three-dimensional Euclidean space and on the two- and 
three-dimensional sphere and hyperboloid. Finally, in \cite{GRSh} a Table of 
Path Integrals will be accomplished which attempts to summarize the 
achievements of solving Feynman path integrals in general, i.e., this will
include not only the numerous applications of the Basic Path Integrals, like
general quadratic Lagrangians with electric and magnetic fields, Coulomb
potentials, monopole systems and path integral formulations for group spaces,
but will also contain path integral formulations for boundary conditions, 
point interactions, coherent states, fermions, supersymmetric quantum 
mechanics, and some field theory formul\ae, respectively the generating 
functionals.

All the known solutions of the Basic Path Integrals are related to the quantum
motion on the $SU(2)$--sphere and $\SU(1,1)$--hyperboloid, respectively their
symmetry group \cite{INJUa}. The quantum motion can have a discrete, or a
continuous spectrum, or both, and the relevant spectrum we need emerging from 
the spectrum of the group manifold $\SU(1,1)$ is of the form (by $m$ I denote
the mass of a test particle, $p>0$ is its momentum)
\begin{equation}
  E_{\sigma,j_0}=-\hbarm[j_0^2+\sigma(\sigma+2)]\enspace,\qquad
  \begin{array}{ll}
  &j_0=0,\sigma=-1+\i p         \enspace,  \\
  &j_0=2n\ (n\in\bbbn),\sigma=-1\enspace,  \end{array}
\end{equation}
for the continuous, respectively discrete spectrum. In this respect most of 
the usual solvable problems, free motion and potential problems, are covered 
within this scheme, including the four Basic Path Integrals.

Looking at the number of coordinate system representations on the 
$\SU(1,1)$--hyperboloid, respectively the $\OO(2,2)$--hyperboloid \cite{KAMIj} 
one realizes that all the known solutions in terms of cartesian, spherical or 
parabolic coordinates come from just a few special coordinate space 
representations of the corresponding matrix elements of the group. This means 
that the power of the majority of the remaining coordinate space 
representations has not been exploited yet, with the exceptions of elliptic 
and spheroidal coordinates in flat space \cite{GROad}, and elliptic and 
elliptic--cylindrical coordinates on the sphere \cite{GKPSa}. The main 
difficulty one encounters in the investigation of the more complicated, i.e., 
parametric coordinate systems, is that very little seems to be known about 
the corresponding theory of special functions in terms of these coordinates. 
These special functions are usually defined by means of recurrence relations. 
These recurrence relations usually cannot be resolved in terms of well-defined 
power series as it is the case for the (confluent) hypergeometric functions.

Concerning path integrals, one is interested in expansion theorems like
(\ref{numAb}) or (\ref{numAa}). The former corresponds to an expansion
connecting two solutions of the Schr\"odinger equation which contains no
free parameter, the second connecting two solutions of the Schr\"odinger 
equation which contains one parameter, which defines the semi-axis of the
ellipse. Usually, such expansion theorems can be obtained by considering the
overlap functions between the wave-functions of two coordinate systems, i.e.,
one considers an interbases expansion theorem (\ref{numAc}). A considerable
effort has been done to investigate such interbases expansion for potential
problems connecting the well-known spherical and parabolic coordinate systems, 
e.g.~\cite{KMPa} (for a more comprehensive bibliography, e.g.\ 
\cite{GROad,GROPOd}), because the corresponding solutions of the Schr\"odinger
equation take on a well-known form. The difficulties arise if one considers
the relevant expansions of say spherical coordinates, and a parametric 
coordinate system, say, spheroidal coordinates. Here it is often necessary
to construct the solutions of the Schr\"odinger equation in terms of the
parametric coordinate systems by means of the interbases expansion
coefficients from an already known solution. In this procedure one uses
the corresponding observables of the parametric coordinate system, and this 
approach guarantees at the same time that the so constructed solutions are 
properly normalized; e.g., the result of \cite{GKPSa}, where this issue was 
addressed for the spherical and cylindrical coordinate systems on the sphere 
$S^{(3)}$ to construct the solution of the Schr\"odinger equation of the two 
one-parametric ellipso-cylindrical coordinate systems.

The question of taking into account two-parametric coordinate systems is even
more difficult, and in comparison to one-parametric systems even less is 
known. For instance, the investigations of the harmonic oscillator in 
ellipsoidal coordinates \cite{KLPS}, and the quantum motion on the 
three-dimensional sphere in terms of ellipsoidal coordinates \cite{AKPSZ}.

The notion of the so-called ``hidden symmetry'' or ``dynamical symmetry'' in
an ordinary potential problem stems from the following observation: One 
considers the quantum motion on a group manifold, say on the two-dimensional
hyperboloid $\Lambda^{(2)}$: $\vec u^2=u_0^2-u_1^2-u_2^2=1$. There exists nine 
coordinate systems on $\Lambda^{(2)}$ which separates the Schr\"odinger,
respectively the Helmholtz equation. Choosing one coordinate system, say the
spherical coordinate system, and separating off the circular variable $\vphi
\in[0,2\pi)$ one is left with the Schr\"odinger equation for the potential 
problem $V(\tau)=\hbarm(l^2-1/4)/\sinh^2\tau$, $\tau>0,\ (l\in\bbbz)$. Taking 
hyperbolic coordinates, and separating off the continuous variable $\tau_2\in
\bbbr$ one is left with the Schr\"odinger equation for the potential $V(\tau
_1)=\hbarm(k^2-1/4)/\cosh^2\tau_2$, $\tau_2\in\bbbr,\ (k\in\bbbr)$. Taking 
horicyclic coordinates and separating off the continuous variable $x\in\bbbr$ 
one is left with the Schr\"odinger equation for the Liouville potential $V(
\tau)=\hbarm(k^2-1/4)\e^{2\vrho}$, $\vrho\in\bbbr,\ (k\in\bbbr)$. Therefore 
we see that the dynamical symmetry ``hidden'' in a well-known potential 
problem corresponds in each case to the same quantum motion on the 
hyperboloid, however realized in different coordinate space representations, 
and the symmetry is the rotational symmetry of the hyperboloid $\Lambda^{(2)}
\subset\SO(2,1)/\OO(2)$. Similar, the P\"oschl-Teller potential is related to the 
quantum motion on 
$\SU(2)$, and the modified P\"oschl-Teller potential is related to $\SU(1,1)$. 
Actually, every path integral which is solvable can be reformulated in terms
of a quadratic--form--Hamiltonian, respectively Lagrangian, i.e., by means of
a ``hidden'' or ``dynamical'' symmetry.
\footnote{In the case of the Coulomb problem, this is the $\OO(4)$-group for the
discrete spectrum and the $\OO(3,1)$-group for the continuous spectrum.}
This is the essence of the Duistermaat-Heckman Theorem \cite{DUHE}. In 
these coordinate system representations for generalized rotation groups (no 
free parameters) one obtains after separating off the ``trivial'' motion of 
subgroups coordinates {\em one dimensional potential problems}.

On the other hand, one-parametric coordinate system produce after separating 
off the subgroup coordinate {\em two-dimensional potentials}. For instance, in
the prolate-spheroidal systems in $\bbbr^3$ one obtains after separating off
the circular variable $\vphi\in[0,2\pi)$ the potential (\cite{GROad},
$l\in\bbbz,k\in\bbbr,\mu>0,\nu\in(0,\pi),d>0$)
\begin{equation}
  V(\mu,\nu)=\hbarm k^2d^2(\sinh^2\mu+\sin^2\nu)
   +(l^2-\bviert){\hbar^2\over2md^2}\bigg(
    {1\over\sinh^2\mu}+{1\over\sin^2\nu}\bigg)\enspace,
\end{equation}
and similarly for oblate spheroidal coordinates. Therefore it is a general
feature that pure subgroup coordinates (like spherical coordinates) generate
by means of the separation procedure one-dimensional potential problems with 
the corresponding ``hidden symmetry'' of the motion on the group space, generic
one-parametric coordinate systems generate two-dimensional potential problems 
with the corresponding ``hidden symmetry'' of the motion on the group space, 
etc. 
\footnote{Note that parabolic coordinates are non-parametric generic 
coordinates which generate a two-dimensional radial harmonic oscillator 
potential, of which the path integral is of the Besselian type, and therefore 
leads to already known results. The range of known special functions, for 
instance the Bessel and Whittaker functions, the Legendre and hypergeometric 
functions, or the spheroidal functions allows only a sufficient number of 
indices to cover the two and three-dimensional wave-functions.}
The couplings in the potentials which emerge and may have originally taken on 
only rational values, can be analytically continued to any number, provided 
the problem at hand remains well-defined.
\footnote{For instance, in Besselian and Legendrian Path Integrals the angular 
momentum number $l$ may become purely imaginary, or complex with 
$\Re(l)>-1/2$.}
The structure of the coordinate systems on the $\OO(2,2)$ hyperboloid is thus
as follows:
\begin{enumerate}
\item $\OO(2,2)\supset E(1,1)$: 10 coordinate systems.
\item $\OO(2,2)\supset\OO(2,1)$: 9 coordinate systems.
\item $\OO(2,2)\supset\OO(1,2)$: 9 coordinate systems.
\item 19 semi--split coordinate systems which are one-parametric.
\item 22 non--split coordinate systems which are two-parametric.
\item 3 non--orthogonal coordinate systems.
\end{enumerate}
The majority of all coordinate systems come from {\em non-subgroup
coordinate systems\/} about next to nothing is known, let alone harmonic
analysis in terms of the corresponding eigenfunctions representations in
terms of the bound states wave--functions and the scattering solutions.

As an instructive example, let us consider the prolate-spheroidal coordinate 
system in four-dimensional Euclidean space, where we separate off the circular 
variables and perform a space-time transformation \cite{GROad,GRSh,GRSb,KLEo}
\begin{eqnarray}       & &\!\!\!\!\!\!\!\!\!\!\!
  \pathint\mu\!\!\pathint\nu{d^4\over4}(\sinh^2\mu+\sin^2\nu)
   \sinh2\mu\sin2\nu\prod_{i=1,2}\pathint{\vphi_i}
         \nonumber\\   & &\!\!\!\!\!\!\!\!\!\!\!\times
   \exp\Bigg\{\ih\!\int_{t'}^{t''}\!\Bigg[{m\over2} d^2\Big(
    (\sinh^2\mu+\sin^2\nu)(\dot\mu^2+\dot\nu^2)
+\sinh^2\mu\sin^2\nu\dot\vphi_1^2+\cosh^2\mu\cos^2\nu\dot\vphi_2^2\Big)
         \nonumber\\   & &\!\!\!\!\!\!\!\!\!\!\!
         \qquad\qquad\qquad\qquad
 +{\hbar^2\over2md^2(\sinh^2\mu+\sin^2\nu)}\Bigg({1\over\sin^2\nu\cos^2\nu}
           +{1\over\sinh^2\mu\cosh^2\mu}\Bigg)\Bigg]dt\Bigg\}
         \nonumber\\   & &\!\!\!\!\!\!\!\!\!\!\!
  =\Big(\hbox{${d^2\over4}$}\sinh2\mu'\sinh2\mu''\sin2\nu'\sin2\nu'\sin2\nu''
   \Big)^{-1/2}
         \nonumber\\   & &\!\!\!\!\!\!\!\!\!\!\!\qquad\times
   \sum_{m_{1,2}\in\bbbz}{\e^{\i m_1(\vphi_1''-\vphi_1')
         +\i m_2(\vphi_2''-\vphi_2')}\over4\pi^2}
         \int_{\bbbr}{dE\over2\pi\i}\e^{-\i ET/\hbar}\int_0^\infty ds''
         \nonumber\\   & &\!\!\!\!\!\!\!\!\!\!\!\qquad\times
   \pathint\mu\pathint\nu\exp\Bigg\{\ih\int_0^{s''}\bigg[
   {m\over2}(\dot\mu^2+\dot\nu^2)+Ed^2(\sinh^2\mu+\sin^2\nu)
         \nonumber\\   & &\!\!\!\!\!\!\!\!\!\!\!
         \qquad\qquad\qquad\qquad
  -\hbarm\Bigg({m_1^2-\viert\over\sin^2\nu}+{m_2^2-\viert\over\cos^2
   \nu}+{m_1^2-\viert\over\sinh^2\mu}-{m_2^2-\viert\over\cosh^2\mu}
   \Bigg)\Bigg]ds\Bigg\}\enspace.\qquad
\label{propagator}
\end{eqnarray}
It is obvious that the corresponding path integral solution in terms of the 
wavefunction expansion is clearly a generalization of the three-dimensional 
case. Whereas in three dimensions the spheroidal wavefunctions yield for $d=0$ 
Legendre functions, the spheroidal wavefunctions in four dimensions have 
(modified) P\"oschl-Teller wavefunctions as their degenerations. I did not 
find explicit representations of the corresponding wavefunctions in the 
literature. However, we can propose heuristically such wavefunctions by taking 
into account the theory of \cite{MESCH}. Looking at the prolate spheroidal
coordinate system, we see that in the limit $d\to0$ the spherical system must 
emerge with the cylindrical coordinate system on $S^{(3)}$ as the proper 
subsystem. Therefore it follows that the wavefunctions in the variable $\nu$ 
must be generalization of the P\"oschl-Teller wavefunctions $\Phi_n^{(\alpha,
\beta)}(z)$, and the wavefunctions in the variable $\mu$ must be again 
generalizations of modified Bessel functions $I_\nu(z)$. The proper quantum 
numbers which must be taken into account are $l\in\bbbn_0$ and $m_1,m_2\in
\bbbz$. Hence we propose spheroidal wavefunctions $\ps_l^{(m_1,m_2)}(\cos\nu;
p^2d^2)$ and $S_l^{(m_1,m_2),(1)}(\cosh\mu;pd)$ together with the following
limiting correspondence ($\gamma=pd$)
\begin{eqnarray}
  \ps_l^{(m_1,m_2)}(\cos\nu;\gamma^2)&\stackrel{\gamma\to0}{\simeq}&
  (\sin\nu)^{|m_1|}(\cos\nu)^{|m_1|}P_l^{(m_1,m_2)}(\cos2\nu)\enspace,\\
  S_l^{(m_1,m_2),(1)}(\cosh\mu;\gamma)&\stackrel{\gamma\to0}{\simeq}&
  {2\pi\over pr}J_{l+1}(pr)\enspace.
\end{eqnarray}
We propose consequently the interbasis
expansion (up to phase factors and redefinition of functions)
 \begin{eqnarray}       & &\!\!\!\!\!\!\!\!\!\!\!
  \exp\Big[\i pd(\sinh\mu\sin\vtheta\sin\nu\cos\alpha
         +\cosh\mu\cos\vtheta\cos\nu\cos\beta)\Big]
         \nonumber\\   & &\!\!\!\!\!\!\!\!\!\!\!
  =\sum_{l=0}^\infty\sum_{m_1,m_2\in\bbbz}2(|m_1|!+|m_2|!+2l+1)
     {l!\Gamma(|m_1|+|m_2|+l+1)\over\Gamma(|m_1|+l+1)\Gamma(|m_2|+l+1)}
         \nonumber\\   & &\!\!\!\!\!\!\!\!\!\!\!\qquad\times
  \e^{\i m_1\alpha+\i m_2\beta}S_l^{(m_1,m_2),(1)}(\cosh\mu;\gamma)
  \ps_l^{(m_1,m_2)}(\cos\vtheta;\gamma^2)
  \ps_l^{(m_1,m_2)}(\cos\nu;\gamma^2)\enspace.\qquad
\end{eqnarray}
By means of this expansion the path integration can be done with the 
conjectural result of the path integral (\ref{propagator})
\begin{eqnarray}       & &\!\!\!\!\!\!\!\!\!\!\!
  K(\mu'',\mu',\nu'',\nu',\vphi_1'',\vphi_1',\vphi_2'',\vphi_2';T)
 =\sum_{l=0}^\infty\sum_{m_1,m_2\in\bbbz}{\e^{\i m_1(\vphi_1''-\vphi_1')
       +\i m_2(\vphi_2''-\vphi_2')}\over4\pi^2}\qquad\qquad
         \nonumber\\   & &\!\!\!\!\!\!\!\!\!\!\!\qquad\times
   2(|m_1|!+|m_2|!+2l+1)
     {l!\Gamma(|m_1|+|m_2|+l+1)\over\Gamma(|m_1|+l+1)\Gamma(|m_2|+l+1)}
         \nonumber\\   & &\!\!\!\!\!\!\!\!\!\!\!\qquad\times
  {2\over\pi}\int_0^\infty p^3dp\,S_l^{(m_1,m_2),(1)}(\cosh\mu'';pd)
  S_l^{(m_1,m_2),(1)\,*}(\cosh\mu';pd)
         \nonumber\\   & &\!\!\!\!\!\!\!\!\!\!\!\qquad\times
  \ps_l^{(m_1,m_2)\,*}(\cos\nu';p^2d^2)
  \ps_l^{(m_1,m_2)}(\cos\nu'';p^2d^2)\e^{-\i\hbar p^2T/2m}\enspace.
\end{eqnarray}
A proper definition of these functions is involved. For instance, the function 
$\ps_\mu^{(m_1,m_2)}$ $(z;\gamma^2)$ must be expanded in terms of $(1-z^2)^
{|m_1|/2}(z)^{|m_1|}P_\mu^{(m_1,m_2)}(2z^2-1)$ with expansion coefficients 
$a_{\mu,\kappa}^{m_1,m_2}(\gamma^2)$, $\kappa\in\bbbz$. These coefficients then 
satisfy some recurrence relations which characterize the spheroidal functions. 
The case of oblate spheroidal coordinates is similar. A detailed study of these 
mathematical issues is beyond the scope of this overview and is not discussed 
here any further, c.f.\ \cite{GROPOd}. Of course, similar considerations can also be 
made for path integral formulations in four-dimensional pseudo-Euclidean space.

Therefore new solutions, expansion theorems, and overlap functions, i.e.,
interbases expansions, for parametric coordinate systems on the $\SU(1,1)$-, 
respectively the $\OO(2,2)$-hyperboloid would provide the possibility to study 
in a more systematic way the following issues:
\begin{enumerate}
\item New special functions will provide new identities between (old and new) 
      special functions.
\item Expansion theorems of plane waves into parametric waves. The relations
      between other wave-expansions then can be obtained by the corresponding
      overlap functions.
\item The solution of the Schr\"odinger equation in parametric coordinate
      systems in terms of parametric special functions, in flat space and
      spaces of (positive and negative) constant curvature alike in terms
      of Lam\'e and Heun-functions and polynomials \cite{KAMIv,PAWI}.
\item The investigation of old potential problems in new settings.
\item The analysis of physical systems possessing symmetry properties which can
      be expressed in parametric coordinates as, e.g., in the two-center 
      problem.
\item The investigation of unfamiliar perturbations for, e.g., the Coulomb
      potential or the harmonic oscillator.
\item Finding new Basic Path Integrals which go beyond the standard solutions.
\end{enumerate}
Exploiting the full power of the group theoretical methods, i.e., symmetry 
methods in the spirit of A.Yu.Smorodinsky, would therefore enlarge our
knowledge and possibilities of harmonic analysis on hyperboloids, special
functions, separation of variables, and, of course, of path integrals.

\section*{Acknowledegement}
I would like to thank the organizers of the conference ``Symmetry Methods
in \hbox{Physics}'', in particular G.Pogosyan and G.Sandukovs\-kaya, for the 
nice atmosphere and the warm hospitality during my stay in Dubna.
\newline
Financial support from the Heisenberg--Landau program is greatfully 
acknowledged.


\input cyracc.def
\font\tencyr=wncyr10
\font\tenitcyr=wncyi10
\font\tencpcyr=wncysc10
\def\cyrrm{\tencyr\cyracc}
\def\cyrit{\tenitcyr\cyracc}
\def\cyrcp{\tencpcyr\cyracc}


\end{document}